\documentclass[10pt]{IEEEtran}
\usepackage{array}
\usepackage{graphicx, subfigure}
\usepackage{float}
\usepackage{bm}
\usepackage{subfloat}
\usepackage{multirow}
\usepackage{rotating}
\usepackage{epstopdf}
\usepackage{amsmath} 
\usepackage{amssymb}  
\usepackage{color}
\usepackage{algpseudocode}
\usepackage{algorithm}

\begin{document}

\bibliographystyle{plain}

\title{Proportional fairness in wireless powered CSMA/CA based IoT networks}

\author{Xiaomin Chen\IEEEauthorrefmark{1}, Zhan Shu\IEEEauthorrefmark{2}, Kezhi Wang\IEEEauthorrefmark{1}, Fangmin Xu\IEEEauthorrefmark{3} and Yue Cao\IEEEauthorrefmark{1}\\
\IEEEauthorrefmark{1} Department of Computer and Information Sciences, Northumbria University, UK \\
\IEEEauthorrefmark{2} Faculty of Engineering and the Environment, University of Southampton, UK\\
\IEEEauthorrefmark{3} Key Laboratory of Universal Wireless Communications, Beijing University of Posts and Communications, China}

\maketitle
\thispagestyle{empty}

\begin{abstract}
This paper considers the deployment of a hybrid wireless data/power access point in an 802.11-based wireless powered IoT network. The proportionally fair allocation of throughputs across IoT nodes is considered under the constraints of energy neutrality and CPU capability for each device. The joint optimization of wireless powering and data communication resources takes the CSMA/CA random channel access features, e.g. the backoff procedure, collisions, protocol overhead into account. Numerical results show that the optimized solution can effectively balance individual throughput across nodes, and meanwhile proportionally maximize the overall sum throughput under energy constraints. 
\end{abstract}

\section{Introduction}
 In an IoT paradigm, smart objects with sensory, computation and communication capabilities are
distributed throughout the environment, and inter-connected via the Internet to conduct various masks, e.g. environmental monitoring, asset tracking, smart homes, smart cities and so on. The IoT concept could lead to unlimited possibilities in all aspects of our daily lives.
However the means to power IoT devices becomes a big stumbling block on the way to the mass adoption of
the paradigm, simply because of the cost, inconvenience, or in certain circumstances infeasibility of wiring remote nodes or constantly
recharging or replacing batteries. 

The recent progress in Wireless Power Transfer (WPT) and Energy Harvesting (EH)~\cite{EH, WPT}
technologies have led to the emergence of Wireless Powered Communication Networks (WPCNs)~\cite{WPCN} in which communication nodes are powered by wireless power transmitters. A typical model of WPCNs contains a Hybrid Access Point (H-AP) and a group of off-grid nodes. The H-AP serves as both a conventional wireless AP and a wireless power source.  Wireless power is transfered to the end users in the downlink, and the users utilize the harvested energy to transmit information to the H-AP in the uplink.  The joint allocation of wireless power and information
communication resources in such a network has been studied e.g. in~\cite{276, 277, 278, 279}. The majority of current work schedules uplink information transmission in a simple TDMA fashion. This scheduling strategy simplifies the system by assigning each user with a dedicated slot for data transmission in a pre-determined order. Optimal allocation of resources, such as the time partition between downlink power transfer and uplink data transmission, transmit power, smart antenna parameters, is derived accordingly. However, the results provides little insight when integrating WPT technologies into existing wireless access networks, such as 802.11 WLANs and 802.15.4 LR-WPANs, as these networks manage channel access in terms of Carrier Sense Multiple Access/Collision Avoidance (CSMA/CA) rather than TDMA. Different from pre-determined scheduling policies (such as TDMA), in a CSMA/CA based network, nodes contend for access to the wireless medium. The allocation of wireless powering durations across nodes in CSMA/CA based WPCNs should take the random backoff procedure, collisions and other MAC layer protocol overhead into account.

In this paper, we consider the deployment of a H-AP in an 802.11-based wireless powered IoT network.  IoT nodes carry out duty-cycled monitoring tasks and get charged by RF beamforming from the H-AP. A harvest-then-transmit strategy is employed for scheduling downlink RF power and uplink information transmission.  Data transmission is based on 802.11 Power Save Mode (PSM). The proportionally fair allocation of throughputs across IoT nodes is derived under the constraints of energy neutrality and CPU capability for each device. The non-convex utility optimization problem is solved by using the Block Coordinate Descent algorithm. In each coordinate direction, the original problem can be converted into a DC programming problem and solved using the standard DC iterative algorithm.  The optimized allocation allows the trade-off of throughput/RF charging duration/air-time to be performed in a principled manner. The optimized number of samples in each cycle equals to the maximum number that its CPU can process concurrently. The throughput for each individual node is effectively balanced by lowering down the air-time for a more aggressive node, and on the contrary, boosting it up for a less aggressive one. The air-time compromises when the node's operation is constrained by the amount of energy available to use. To the best of our knowledge, this is the first work that the resource allocation of wireless powering and data communication is considered together with CSMA/CA based random channel access features. Our findings provide meaningful insight on the implementation of RF powering technology in WiFi-based IoT networks.

\section{System Model}

We consider a star topology IoT network consisting of a hybrid energy and information AP and a set of IoT nodes $\mathcal{N}$ with $\left|\mathcal{N}\right|=N$.  
The IoT nodes are duty-cycled monitoring devices following a common operational pattern: data is acquired from the surrounding environment by a sensor module, processed by a controller and then sent to the network by a Wireless Network Interface Controller (WNIC). The wireless interface is an 802.11 WLAN (WiFi) module. Due to the shortcomings of limited communication range and high power consumption, WiFi is not an appropriate option for IoT connectivity.  However as in-building WiFi coverage is now almost ubiquitous, the latest IEEE 802.11ah/ax standards and the development of low power WiFi chipsets make WiFi a handy and cheap IoT connectivity option. 
 
In a time slotted WLAN channel, the process of data acquisition, processing and transmission repeats across time slots. We define each repetitive process as a cycle. To reduce power consumption, the WLAN interface operates in the 802.11 PSM~\cite{PSM}. During the phases of data sensing and processing, the WNIC stays in sleep mode. After the acquired raw data is processed and encapsulated into the packet format, the WNIC wakes up and starts the 802.11 Distributed Coordination Function (DCF) random channel access process. Once a transmission opportunity is obtained, the queued packets are sent out over the wireless channel to the H-AP. Since the majority of traffic in a typical IoT network is uplink, we neglect downlink traffic in our analysis. The information channel between the H-AP and end nodes is assumed to be error-free, i.e. packet errors are only caused by collisions. 

The IoT nodes are off-grid nodes equipped with an energy harvesting module and a rechargeable battery. The H-AP is equipped with multiple antennas. For each individual HAP-node pair, a harvest-then-transmit strategy is employed in a half-duplex channel. Each node is charged by the H-AP point-to-point via RF energy beamforming~\cite{Beamforming} when its wireless interface is sleeping. The wireless charging process is terminated when the wireless interface wakes up for data transmission.  To maintain unattended nodes to ideally last for an unlimited period of time, the energy neutrality principle is considered~\cite{Neutrality}, that is, the consumed energy should be, at most, equal to the harvested energy from the H-AP over the long term. It is assumed that the H-AP can emit multiple RF beams to distinct nodes simultaneously by using multiple antenna space division multiple access (SDMA) technology~\cite{SDMA}. The transmit power in each spatial stream can be independently tuned in terms of individual channel quality. The interference across spatial streams is neglected in this work.

\section{MAC analytical model}

The channel access is managed by 802.11 CSMA/CA scheme when the WLAN NIC is in active mode. When a node senses the channel idle for a DCF IntreFrame Space (DIFS) period, it waits for an additional random number of time slots before sending the data into the channel. The random number is selected uniformly between $0$ and the minimum contention window $W_i$. In this work we disable retransmissions as the optimal $W_i$ will be derived to proportionally maximize the network throughput, the collision probability can thus be tuned to the optimum to achieve the desirable performance right after the first attempt.  

\subsection{ Per-node Markov chain}

\begin{figure}
 \centering
  \includegraphics[width=0.85\columnwidth,height=3.5cm]{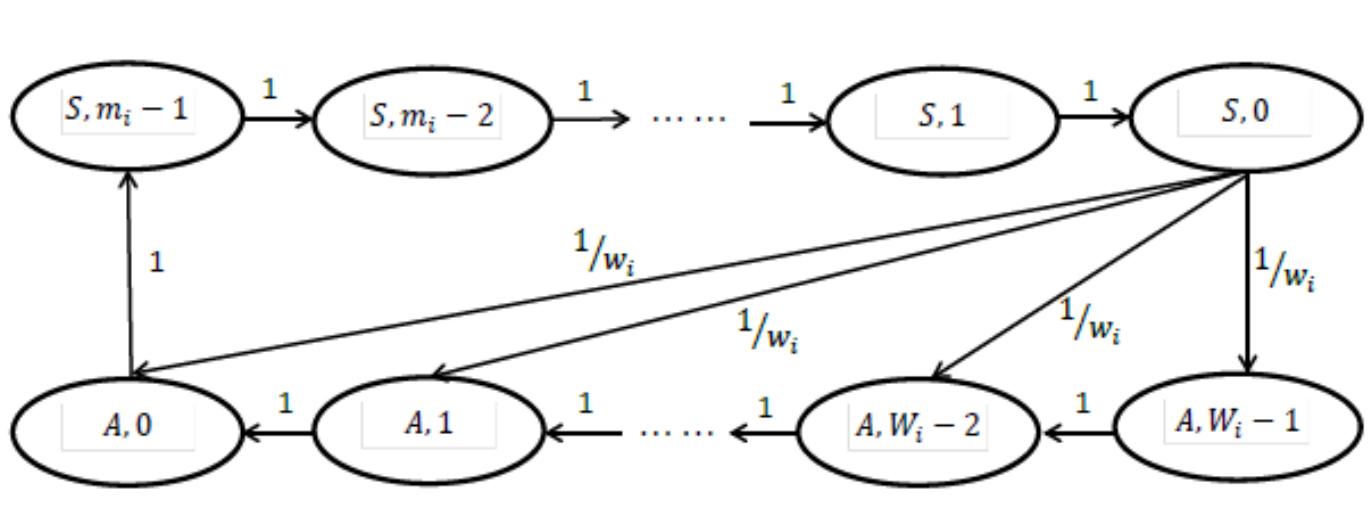}
  \caption{Markov chain for duty-cycled monitoring IoT nodes in 802.11 PSM}\label{fig:MarkovChain}
\end{figure}

The MAC behavior of the considered network can be modeled using a Markov chain as depicted in Fig.~\ref{fig:MarkovChain}. 
State $(A, k)$ denotes that the wireless interface is in active mode and the size of the backoff counter is $k$. The initial backoff counter is uniformly chosen between ${[0,W_i-1]}$.  As no retransmissions are considered, the backoff counter size has only $W_i$ possible values. While the medium is sensed to be idle at the beginning of a time slot, the backoff counter is decremented by one. Transmission is attempted when $k=0$. State $(S, k)$ represents that the wireless interface is now in sleep mode and the number of time slots to go before the wake-up time is $k$.  For node $i$, the number of time slots during which the wireless interface is sleeping is $m_i$. The number of sleeping slots decrements by one after each slot. The wireless interface wakes up after state $(S,0)$ which is followed by state $(A,k),  k=\{0,1,\cdots,W_i-1\}$ with the probability $1/W_i$. The transition probability is given by
\begin{equation*}
    P{[\ (A,k) \ |\ (S,0)\ ]}=\frac{1}{W_i},  \ \ \  k=0,1,\dots,W_i-1.
\end{equation*}
In the active mode, the transition probability from state $(A,k)$ to state $(A,k-1)$ is given by
\begin{equation*}
    P{[\ (A,k-1) \ |\ (A,k)\ ]}=1,  \ \ \  k=1, 2, \dots,W_i-1.
\end{equation*}
In the sleep mode, the transition probability from state $(S,k)$ to state $(S,k-1)$ is then
\begin{equation*}
    P{[\ (S,k-1) \ |\ (S,k)\ ]}=1,  \ \ \  k=1, 2, \dots,m_i-1.
\end{equation*}
The transition probability from active mode to sleep mode is given by
\begin{equation*}
    P{[\ (A,0) \ |\ (S,m_i-1)\ ]}=1.
\end{equation*}

Let $\bm{b}$ denote the state stationary distribution in the Markov chain, we have
\begin{equation}\label{Eqn:Markov}
\sum_{k=0}^{W_i-1}b_{A,k}+\sum_{k=0}^{m_i-1}b_{S,k}=1
\end{equation} 
in which
\begin{equation*}
    b_{A,k}=\frac{W_i-k}{W_i}b_{A,0}, \ \ \  k=1,2,\dots,W_i-1
\end{equation*}
and 
\begin{equation*}
    b_{S,k}=b_{A,0}, \ \ \  k=0,1,\dots,m_i-1
\end{equation*}
Solving Eqn.~(\ref{Eqn:Markov}), we obtain
\begin{equation*}
    b_{A,0}=\frac{2}{W_i+2m_i+1}.
\end{equation*}

The main interesting quantity in this MAC analytical model is the node attempt probability $\tau_i$, i.e. the probability that an IoT node attempts to make a transmission at a time slot. A node makes a transmission only if the backoff counter reaches 0 in the active mode,  and thus
\begin{equation}\label{Eqn:tau}
    \tau_i=b_{A,0}=\frac{2}{W_i+2m_i+1}.
\end{equation}

It can be seen that the derived analytical model is a special case of the classic Bianchi model~\cite{Bianchi} with $CW_{max}=CW_{min}$ plus additional `sleep' states.  

\subsection{Per-node throughput}

Each state in the Markov chain could be occupied by a successful transmission, a collision or an idle slot. Let $P^{idle}$ denote the probability that there is no transmission in a time slot, we have
$P^{idle}=\prod_{i=0}^{N-1}(1-\tau_i)$
The probability that node $i$ makes a successful transmission is 
$P_i^{succ}=\tau_i\prod_{j=0, j \neq i}^{N-1}(1-\tau_j)$
The collision probability in a time slot is then
$P^{col}=1-P^{idle}-\sum_i^{N-1}P_i^{succ}$

For IoT node $i$, it is assumed that the number of samples taken during the data acquisition phase in each cycle is $n_i$, and the sampling period is $h_i$ slots. The duration of data acquisition process is thus $n_ih_i$ time slots. After data acquisition, the raw data is processed by the CPU. Modern processors offer multiple cores to perform multitasking, so we assume that data processing is finished within $g_i$ slots, and the maximum samples it can process concurrently is $\bar{n}_i$. The duration of sleep mode can hence be represented as
\begin{equation}\label{Eqn:m}
m_i=n_ih_i+g_i
\end{equation}
As $n_i \geq 1$, $h_i \geq 1$ and $g_i \geq 1$, $m_i$ is greater than 2. 

Each acquired sample is encapsulated into a packet with length of $l_i$ bits. As data packets in monitoring applications are normally short packets, 802.11n Aggregate MAC Service Data Unit (A-MSDU) is used to improve MAC throughputs, and meanwhile reduce energy expenditure~\cite{80211n}. To further save power and make fast recovery from collisions, the Request To Send/Clear To Send (RTS/CTS) handshake scheme is employed before an A-MSDU packet. The transmission duration of an A-MSDU packet is given by
\begin{equation*}
T_i^{A\text{-}MSDU}=T_i^o+n_i\left(\frac{l_i}{R_i}+T_i^{shdr}\right)
\end{equation*}
in which $T_i^o=T^{PHY\text{-}HDR}+T_i^{MAC\text{-}HDR}+T_i^{FCS}$; $R_i$ is the PHY data rate; $T_i^{shdr}$ is the sub-header for each sub-frame within the A-MSDU. The total duration of a successful A-MSDU transmission including the RTS/CTS handshake and ACK is then given by
\begin{equation*}
T_i^{succ}=T_i^{oo}+n_i\cdot\left(\frac{l_i}{R_i}+T_i^{shdr}\right)
\end{equation*}
in which $T_i^{oo}=T_i^o+T^{RTS}+T^{CTS}+3T^{SIFS}+T^{ACK}$
is the protocol overhead associated with a successful A-MSDU transmission.

A collision can be considered to have happened if a CTS does not arrive within a timeout period $T^{tout}$ after a RTS is sent out. The duration of a collision is then
$T^{col}=T^{RTS}+T^{tout}$, in which 
$T^{tout}=T^{SIFS}+T^{CTS}+\sigma$ and $\sigma$ represents the duration of a physical time slot.

The throughput of node $i$ is therefore given by
\begin{equation*}
S_i=\frac{n_il_iP_i^{succ}}{P^{idle}\sigma+\sum\limits_{j=0}^{N-1}P_j^{succ}T_j^{succ}+P^{col}T^{col}}
\end{equation*}
The throughput expression can be reformulated by letting $\alpha_i=\tau_i/(1-\tau_i)$ as
\begin{equation}\label{Eqn:thrpt}
S_i=\frac{\alpha_in_il_i}{XT^{col}}
\end{equation}
in which 
\begin{equation*}
\begin{aligned}
&X=\frac{\sigma}{T^{col}}+\sum\limits_{j=0}^{N-1}\frac{\frac{l_j}{R_j}+T_j^{shdr}}{T^{col}}n_j\alpha_j+\sum\limits_{j=0}^{N-1}(\frac{T_j^{oo}}{T^{col}}-1)\alpha_j\\
&\ \ \ \ +\prod_{j=0}^{N-1}(1+\alpha_j)-1
\end{aligned}
\end{equation*}

\section{Energy consumption model}

In this section, we will provide an energy consumption model for duty cycled monitoring nodes in a WLAN.  Based on the assumptions, the energy required in each cycle can be partitioned into three parts: for data sensing or acquisition $E_i^{A}$, for data processing $E_i^P$, and data transmission $E_i^T$. Additionally, a small fraction of energy is consumed to run background tasks, such as running the operating system, switching between wake-up and sleep modes, clock synchronizing and so on.  The total system energy consumed in each cycle is therefore
\begin{equation}\label{Eqn:energy}
E_i=E_i^{A}+E_i^{P}+E_i^{T}+E_i^{BG}
\end{equation}
Next we will derive the expression for $E_i^{A}$, $E_i^{P}$ and $E_i^{T}$ respectively. $E_i^{BG}$ is considered as a constant value in our analysis.

\subsection{Energy consumed for data acquisition and processing}

The energy drained from the battery for data acquisition and processing is linear with the number of samples taken in each cycle given the assumption that each sampling and processing operation costs a constant amount of energy. Assuming that the power for acquiring samples from environment is $P_i^{A}$, and each sampling operation only takes 1 slot, the energy consumed for data acquisition in each cycle is given by
$E_i^{A}=n_iP_i^A\sigma$
The energy consumed for processing $n_i$ samples is then 
$E_i^P=n_iP_i^Pg_i\sigma$e
given that the power for processing a sample is $P_i^{P}$.

\subsection{Energy consumed for data transmission}

In the process of data transmission, the WNIC performs 802.11 DCF backoff first before transmitting an A-MSDU packet. It is indicated in~\cite{WNICenergy} that the power consumption during data transmission can be classified into four levels depending on the operation the WNIC performs: i) listening to the channel, $P_i^{L}$; ii) receiving/overhearing traffic, $P_i^{R}$; iii) transmitting traffic, $P_i^{T}$ and vi) background tasking $P_i^{BG}$. In the following we will respectively calculate the energy consumption during the backoff stage and the data transmission stage. 

\subsubsection{Energy consumption during DCF backoff}
Energy is mainly consumed for listening to the channel (i.e. physical carrier sensing) during the backoff procedure. 
Once packets are available in the buffer, the wireless interface starts to listen to the channel. After the channel is sensed as idle for a DIFS, it sets the backoff counter and starts to count down, during which it keeps listening to detect any transmissions from other nodes. Due to the deployment of RTS/CTS scheme, if the countdown process is suspended by an ongoing transmission from others, the wireless interface switches to \emph{virtual} carrier sensing by setting the Network Allocation Vector (NAV) with the duration of the interrupting data transmission. That is, it defers listening to the channel until the interrupting transmission is finished. The energy consumed during this process can thus be neglected. As there are no retransmissions allowed in our setting, the average number of time slots during which node $i$ listens to the channel is $(W_i-1)/2$. The energy consumed during the backoff process is thus 
\begin{equation*}
E_i^{BO}=\left(T^{DIFS}+\frac{W_i-1}{2}\sigma\right)P_i^L.
\end{equation*}

\subsubsection{Energy consumed for transmitting the data packet}

After the DCF backoff process, node $i$ gets access to the channel and starts the transmission of the A-MSDU data packet. 
An A-MSDU transmission could succeed or end up with a collision.
The energy required to transmit an A-MSDU is
\begin{equation*}
\begin{aligned}
\epsilon_i^{A\text{-}MSDU}=&(T^{RTS}+T_i^{A\text{-}MSDU})P_i^T+(T^{CTS}+T^{ACK})P_i^R\\&+2T^{SIFS}P_i^L
\end{aligned}
\end{equation*}
The energy consumed if a collision occurs is
\begin{equation*}
\epsilon_i^{col}=T^{RTS}P_i^T+T^{tout}P_i^{L}
\end{equation*}
Taking into account the two possible consequences, the expected energy that node $i$ consumes for transmitting the aggregated data packet after the DCF backoff is given by
\begin{equation*}
\begin{aligned}
E_i^{DATA}=&\prod_{\substack{j=0 \\ j\neq i}}^{N-1}(1-\tau_j) \epsilon_i^{A\text{-}MSDU}
                   +\Big(1-\prod_{\substack{j=0 \\ j\neq i}}^{N-1}(1-\tau_j)\Big) \epsilon_i^{col}
\end{aligned}
\end{equation*}
The total energy consumption throughout the entire data transmission process is therefore given by
\begin{equation*}
E_i^T=E_i^{BO}+E_i^{DATA}.
\end{equation*}

\section{Proportional fairness}

One of the most significant performance metrics in IoT networks is throughput. In the considered network, the resources, such as bandwidth, airtime, are shared by all of the participating nodes, the increase of throughput for one node comes at the cost of the decrease for others.  It can be seen from Eqn.~(\ref{Eqn:thrpt}) that given a fixed packet size $l_i$ and a fixed PHY data rate $R_i$, the throughput for each individual node is jointly determined by the number of samples $\bm{n}:=(n_0 ,n_1,\cdots,n_{N-1})$ taken in each cycle and the attempt probability parameter $\bm{\alpha}:=(\alpha_0,\alpha_1,\cdots,\alpha_{N-1})$ of all the nodes in the network.
In this section we will derive the optimal number of samples taken in each cycle as well as the optimal node attempt probability for each individual node to maximum the total network throughput in a proportionally fair way.

\subsection{Network utility optimization problem}
The network utility function is defined as the sum of the log of node throughputs. The utility optimization problem is to obtain optimum $\bm{\alpha}:=(\alpha_0,\alpha_1,\cdots,\alpha_{N-1})$ and $\bm{n}:=(n_0,n_1,\cdots,n_{N-1})$ to maximize the utility function subject to the constraints on the node computation capability and energy neutrality.
\begin{align}\tiny
&\max\limits_{\bm{n},\bm{\alpha}} \ U(\bm{n},\bm{\alpha}):=\sum_{i=0}^{N-1}\log S_i(\bm{n},\bm{\alpha}) \label{Eqn:utility} \\
&s.t. \ \ \ \ 0<\alpha_i  \leq 0.5 \ \ \ \ \ \ \ \ \ \ \ i=0,1,\dots,N-1,\label{Cst1} \\ 
&\ \ \ \ \ \ \ \  1 \leq n_i \leq \bar{n}_i \ \ \ \ \ \ \ \ \ \ \ \ i=0,1,\dots,N-1,\label{Cst2} \\
&\ \ \ \ \ \ \ \ E_i (n_i, \bm{\tau}) \leq \phi_im_i\sigma \ \ \ \  i=0,1,\dots,N-1. \label{Cst3} 
\end{align}
As $W_i \geq 1$ and $m_i \geq 2$, it follows that $\tau_i \leq 1/3$ according to Eqn.~(\ref{Eqn:tau}), and thus $0 \leq\alpha_i\leq0.5$. Equation~(\ref{Cst2}) enforces the constraint on the maximum number of samples that the CPU of node $i$  can process concurrently within $g_i$ time slots. Constraint~(\ref{Cst3}) ensures that the average energy consumed in each cycle, given by Eqn.~(\ref{Eqn:energy}), does not exceed the RF power supplied by the H-AP, in which $\phi_i$ is the RF power received at node $i$, and $m_i$ is the number of slots allocated for wireless powering, given by Eqn.~(\ref{Eqn:m}).  By plugging the expressions for $E_i^A$, $E_i^P$ and $E_i^T$ into Eqn.~(\ref{Eqn:energy}), Constraint~(\ref{Cst3}) can be further expanded as
\begin{equation}\label{Eqn:energy1}
\begin{aligned}
A_i n_i+\frac{B_i}{\alpha_i}+C_in_i\prod_{\substack{j=0\\j \neq i}}^{N-1}\frac{1}{1+\alpha_j}+D_i\prod_{\substack{j=0\\ j \neq i}}^{N-1}\frac{1}{1+\alpha_j}\leq F_i
\end{aligned}
\end{equation}
in which 
\begin{equation*}
\begin{aligned}\tiny
&A_i=\epsilon_i^A+\epsilon_i^P-(\phi_i+P_i^L)h_i\sigma \\ 
&B_i=\sigma P_i^L \\
&C_i=\left(\frac{l_i}{R_i}+T_i^{shdr}\right)P_i^T \\
&D_i=(T^{CTS}+T^{ACK})P_i^R+(2T^{SIFS}-T^{tout})P_i^L+T_i^oP_i^T\\
&F_i=\phi_ig_i\sigma-E^{BG}-(T^{DIFS}+T^{tout}-g_i\sigma)P_i^L-T^{RTS}P_i^T
\end{aligned}
\end{equation*}

\subsection{Solving the non-convex optimization problem}

It can be verified by inspection of Hessian matrix that the objective utility function is not jointly concave in $\bm{n}$ and $\bm{\alpha}$, and constraint~(\ref{Eqn:energy1}) is not convex either. The problem does not conform to a standard convex optimization problem. To solve this problem, we proceed by finding that given a fixed $\bm{\alpha}$, the original problem is a DC programming problem in $\bm{n}$, and likewise, the problem is DC in $\alpha_i$ given a fixed $\bm{n}$ and $\{\alpha_j\}_{j \neq i}$. 

\subsubsection{DC programming in the coordinate direction of $\bm{n}$}
Given a fixed $\bm{\alpha}$, the original optimization is transformed as below by removing constant terms, 
\begin{align}\tiny
& \max\limits_{\bm{n}} \  f_1:=\sum_{i=0}^{N-1}\log n_i-N\log X (\bm{n}) \label{Eqn:prbm1}\\
s.t. &\nonumber\\
&1 \leq n_i \leq \bar{n}_i,  \ \ \ \ i=0,1,\dots,N-1\label{Cst4}   \\
&\big(A_i+C_i\prod_{\substack{j=0\\j \neq i}}^{N-1}\frac{1}{1+\alpha_j}\big)n_i\leq F_i-\frac{B_i}{\alpha_i}-D_i\prod_{\substack{j=0\\j \neq i}}^{N-1}\frac{1}{1+\alpha_j}, \nonumber\\ 
& \ i=0,1,\dots,N-1. \label{Cst5}
\end{align}
If we define 
\begin{equation*}
s(\bm{n})=\sum_{i=0}^{N-1}\log n_i
\end{equation*}
and 
\begin{equation*}
t(\bm{n})=N\log X(\bm{n})
\end{equation*}
it can be verified by inspecting the second derivatives that both $s(\bm{n})$ and $t(\bm{n})$ are concave in $\bm{n}$. 
The objective function
\begin{equation*}
\max\limits_{\bm{n}}\  f_1(\bm{n}) =s(\bm{n})-t(\bm{n})
\end{equation*}
is thus a Difference of Convex (DC) function~\cite{DC}. As both constraint~(\ref{Cst4}) and constraint~(\ref{Cst5}) are convex as well, the optimization problem is a DC programming problem, which can be solved using the iterative algorithm described in Algorithm~\ref{Alg:DCn} given the fact that $t(\bm{n})$ is continuously differentiable in $\bm{n}$. 
\begin{algorithm}\small
\caption{Iterative algorithm to solve $\bm{n}$ for the DC problem given $\bm{\alpha}$ fixed}\label{Alg:DCn}
\begin{algorithmic}[1]
\State Initialize $\bm{n^{(0)}}$, set k = 0 (iteration number).
\Repeat
\State Define an auxiliary function $\hat{f_1}^{(k)}(\bm{n})$ as 
\begin{equation*}
\hat{f_1}^{(k)}(\bm{n}):=s(\bm{n})-t(\bm{n}^{(k)})-\nabla t(\bm{n}^{(k)})\cdot(\bm{n}-\bm{n}^{(k)})
\end{equation*}
 \Statex \ \ \ \ where $\nabla t(\bm{n}^{(k)})$ is the gradient of $t$ at $\bm{n}^{(k)}$.
\State Solve the convex optimization problem
\begin{equation*}
\bm{n}^{(k+1)}= \arg \max\limits_{\bm{n} \in \mathcal{Z} }\hat{f_1}^{(k)}(\bm{n}) 
\end{equation*}
\Statex \ \ \ \ where $ \mathcal{Z}$ is the convex set defined by constraint~(\ref{Cst4})  and~(\ref{Cst5}) 
\State $k \gets k+1$
\Until {The sequence $\{ f_1(\bm{n}^{(k)})\}$ converges.}
\end{algorithmic}
\end{algorithm}

\subsubsection{DC programming in the coordinate direction of $\alpha_i$}

If we fix $\bm{n}$ and $\{\alpha_j\}_{j \neq i}$, the problem is transformed as 
\begin{align}\tiny
& \max\limits_{\alpha_i} \  f_2:=\sum_{i=0}^{N-1}\log \alpha_i-N\log X (\alpha_i) \label{Eqn:prbm2}\\
s.t. & \nonumber\\
& 0 \leq \alpha_i \leq 0.5, \label{Cst6}   \\
& B_i\alpha_i^{-1} \leq F_i-A_i n_i-(C_in_i+D_i)\prod_{\substack{j=0\\j \neq i}}^{N-1}\frac{1}{1+\alpha_j}, \label{Cst7} \\
& \prod_{\substack{k=0\\ k \neq i, j}}^{N-1}\frac{C_jn_j+D_j}{1+\alpha_k}\cdot(1+\alpha_i)^{-1}\leq F_j-A_jn_j-B_j\alpha_j^{-1}  \nonumber\\
&\ \ \ \ \ \ \ \ \ \ \ \ \ \ \  \ \ \ \ \ \ \ \ \ \ \ \ \ \ \ \ \ \ \ \ \ \ \ \ \ \ \ \ j \in\mathcal{N}, j \neq i. \label{Cst8}  
\end{align}
It can be seen that the objective function $f_2$ is a DC function provided that $p=\sum_{i=0}^{N-1}\log \alpha_i$ and $q=N\log X (\alpha_i)$ are both concave in $\alpha_i$. Also, the three constraints are all convex in $\alpha_i$ given that $B_i>0$ and $\prod_{{k=0, k \neq i, j}}^{N-1}(C_jn_j+D_j)(1+\alpha_k)^{-1}>0$. The iterative  algorithm to solve this DC programming problem is described in Algorithm~\ref{Alg:DCalpha}.
\begin{algorithm}\small
\caption{Iterative algorithm to solve $\alpha_i$ for the DC problem given that $\bm{n}$ and  $\{\alpha_j\}_{j \neq i}$ are fixed}\label{Alg:DCalpha}
\begin{algorithmic}[1]
\State Initialize ${\alpha_i^{(0)}}$, set k = 0 (iteration number).
\Repeat
\State Define an auxiliary function $\hat{f_2}^{(k)}(\alpha_i)$ as 
\begin{equation*}
\hat{f_2}^{(k)}(\alpha_i):=p(\alpha_i)-q(\alpha_i^{(k)})-\nabla q(\alpha_i^{(k)})\cdot(\alpha_i-\alpha_i^{(k)})
\end{equation*}
\Statex \ \ \ \ where $\nabla q(\alpha_i^{(k)})$ is the gradient of $q$ at $\alpha_i^{(k)}$.
\State Solve the convex optimization problem
\begin{equation*}
\alpha_i^{(k+1)}= \arg \max\limits_{\alpha_i \in \mathcal{X} }\hat{f_2}^{(k)}(\alpha_i) 
\end{equation*}
\Statex \ \ \ \ where $ \mathcal{X}$ is the convex set defined by constraint~(\ref{Cst6}), (\ref{Cst7})
\Statex \ \ \ \ and (\ref{Cst8}). 
\State $k \gets k+1$
\Until {The sequence $\{ f_2(\alpha_i^{(k)})\}$ converges.}
\end{algorithmic}
\end{algorithm}

\subsubsection{BCD algorithm}
Based on the above observations and analysis, the original non-convex optimization problem can therefore be solved by employing Block Coordinate Descent (BCD) algorithm in the continuous domain of $\bm{\alpha}$ and $\bm{n}$~\cite{BCD}. The algorithm is given in Algorithm~\ref{Alg:BCD}.
\begin{algorithm}\small
\caption{BCD algorithm to solve for optimized $\bm{n}$ and $\bm{\alpha}$ }\label{Alg:BCD}
\begin{algorithmic}[1]
\State Initialize ${\bm{\alpha}^{(0)}}$, set k = 0 (iteration number).
\Repeat
\State Fix $\bm{\alpha}=\bm{\alpha}^{(k)}$, solve the DC problem~(\ref{Eqn:prbm1}) using Alg.~\ref{Alg:DCn}
\begin {equation*}
\bm{n}^{(k+1)}= \arg \max\limits_{\bm{n} \in \mathcal{Z} }{f_1}(\bm{n})
\end{equation*}
\State Initialize $i \gets 1$
\Repeat
\State Fix $\bm{n}=\bm{n}^{(k+1)}$ and 
$\alpha_{j}=\begin{cases}
                         \alpha_{j}^{(k+1)} \  \forall j < i\\
                         \alpha_{j}^{(k)} \ \ \ \  \forall j > i\\
                    \end{cases}$, solve the 
\Statex \ \ \ \ \ \ \ \ \ DC problem~(\ref{Eqn:prbm2}) using Alg.~\ref{Alg:DCalpha} to update $\alpha_i$
\begin {equation*}
\alpha_i^{(k+1)}= \arg \max\limits_{\alpha_i \in \mathcal{X} }{f_2}(\alpha_i)
\end{equation*}
\State $i \gets i+1$
\Until 
\State $k \gets k+1$
\Until {The network utility sequence $\{ U(\bm{n}^{(k)}, \bm{\alpha}^{(k)})\}$ converges.}
\end{algorithmic}
\end{algorithm}
\section{Simulation Evaluation}
\subsection{Impact of maximum number of samples allowed per cycle}

We first look at the impact of CPU capability, i.e. the maximum number of samples allowed to be taken in each cycle, on the resource allocation. The protocol parameters used in the simulation are listed in Table~\ref{parameters}. An example of a network with 6 IoT nodes is considered. All of the nodes have the same data packet size, PHY data rate,  sampling frequency, consumption powers for data transmitting/receiving, listening to the channel, sample acquisition/processing, and RF power received from the H-AP. The only difference here is the number of samples allowed to be taken per cycle, i.e. $\bar{n}_i$ for node $i$. The example values for these parameters are listed in Table~\ref{Exmp1Parameters}. The ratio among the consumption power for transmitting, receiving and listening to the channel is assumed to be the same as the experimental observation in the work~\cite{WNICenergy}. Note that the parameter values assumed in this example are only for evaluation of the proposed proportional fair allocation strategy. They do not provide any technical advice on practical RF powering implementation. 
\begin{table}\tiny
\caption{MAC protocol parameters}\label{parameters}
\centering
\begin{tabular}{|c|c|c|c|c|c|}
  \hline
  $T^{SIFS}$           & 16$\mu$s      &  $T^{DIFS}$        & 34$\mu$s & $\sigma$  & 9$\mu$s     \\ \hline
  $T^{ACK}$            &    38.67$\mu$s   & $T^{RTS}$            &    46.67$\mu$s  &  $T^{CTS}$            &    38.67$\mu$s  \\ \hline
$T^{PHY-HDR}$         & 20 $\mu$s&  $L^{MAC-HDR}$         & 36 bytes  &  $L^{shdr}$         & 14 bytes \\ \hline
 $L^{FCS}$         & 4 bytes &  & & &\\ \hline
\end{tabular}
\end{table}

\begin{table}\tiny
\caption{Parameter values in the example for evaluating the impact of maximum sample number}\label{Exmp1Parameters}
\centering
\begin{tabular}{|c|c|c|c|c|c|c|c|c|c|}
  \hline
  $P_i^{T}$           & 15mW      &  $P_i^{R}$        & 11.37mW &  $P_i^{L}$   & 10mW & $l_i$         & 50 bytes & $h_i$         & 3      \\ \hline
  $\phi_i$           &    15mW    & $P_i^{A}$      &    5mW  &   $P_i^{P}$            &    6mW &  $R_i$         & 11Mbps &  $g_i$         & 2 \\ \hline
\end{tabular}
\end{table}

\begin{figure} 
 \centering
  \includegraphics[width=0.55\columnwidth,height=3.5cm]{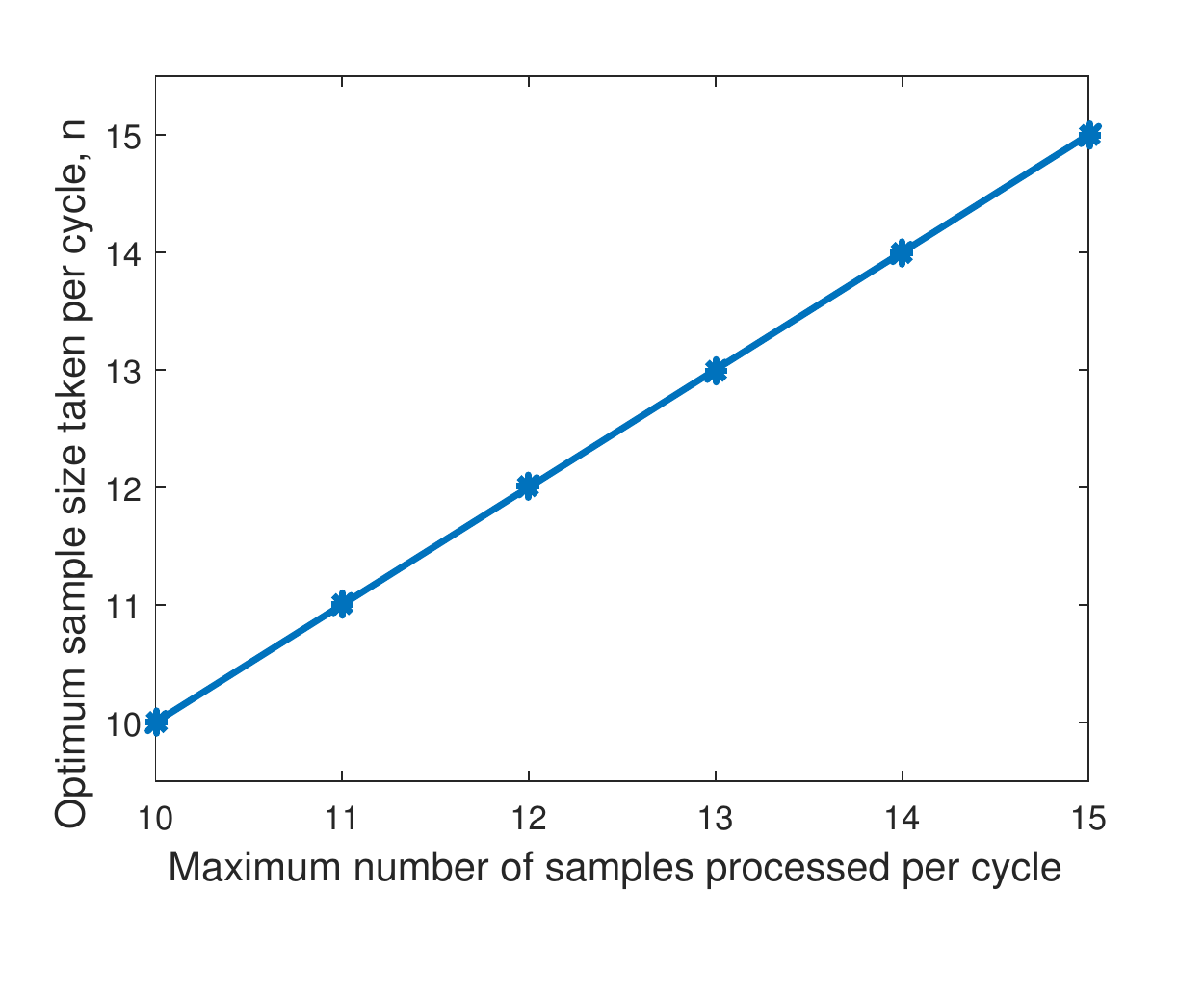}
  \caption{Optimal number of samples per cycle vs the maximum allowed given all other parameters the same for the 6 nodes, parameter values listed in Table~\ref{Exmp1Parameters}}\label{Fig:sampleSizeE1}
\end{figure}

Fig.~\ref{Fig:sampleSizeE1} plots the optimal number of samples taken per cycle versus the maximum number that the CPU can process concurrently for the 6 nodes in the example. It can be seen that for each of the nodes the optimized choice is the maximum, i.e. Constraint~(\ref{Cst2}) is always tight. This result conforms to the objective of proportionally maximizing the sum throughput and the proposed scheduling strategies for wireless charging and data transmission. As all of the samples taken have to be sent out  within each cycle, and the amount of energy charged from the H-AP is linear with the number of samples taken, more samples lead to higher throughput and longer charging duration, so each node chooses the largest possible number.  

\begin{figure}
\centering \subfigure[Energy consumed and received]{
  \includegraphics[width=0.46\columnwidth, height=3cm]{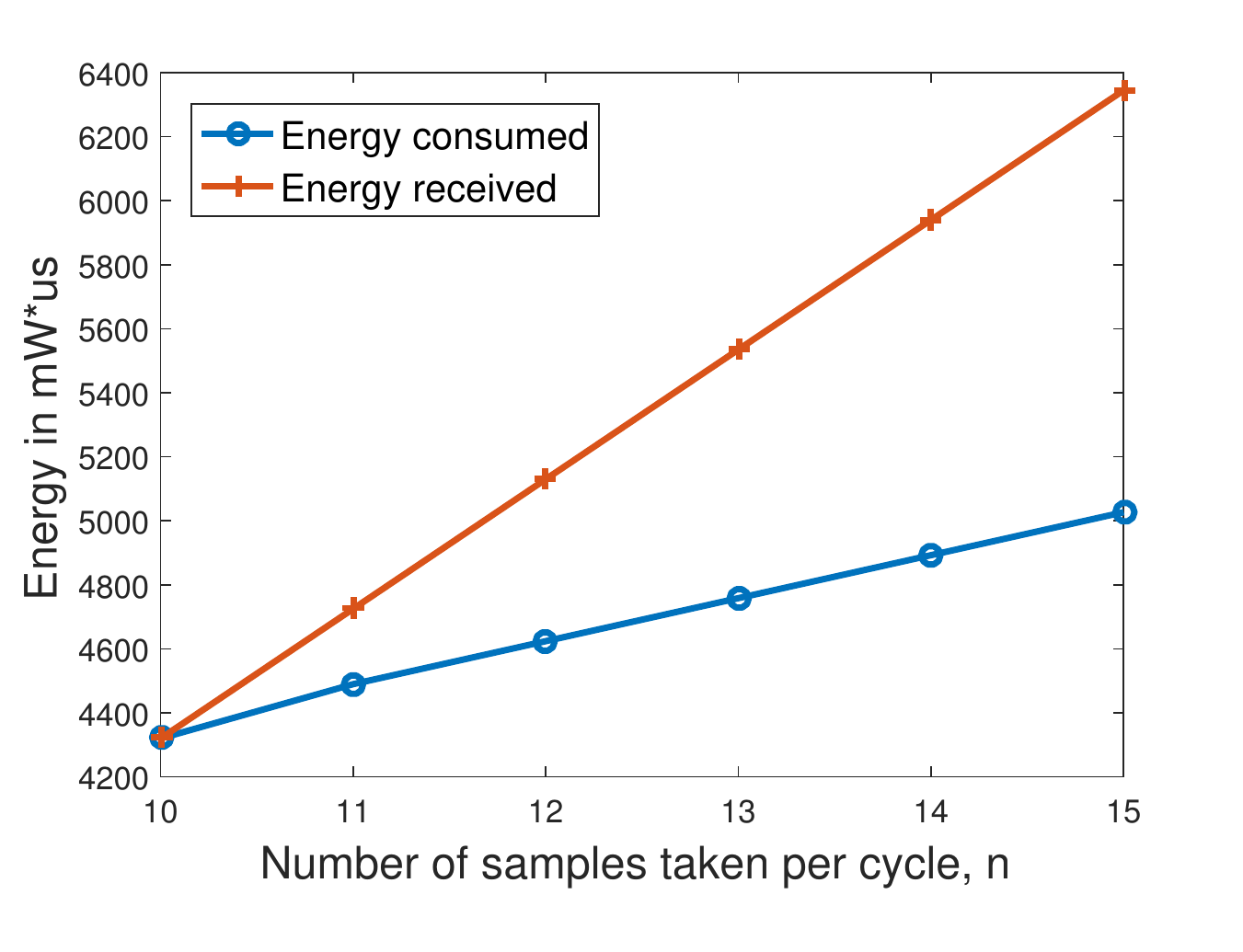}\label{Fig:energyE1}
 }
 \subfigure[Air-time]{
   \includegraphics[width=0.46\columnwidth, height=3cm]{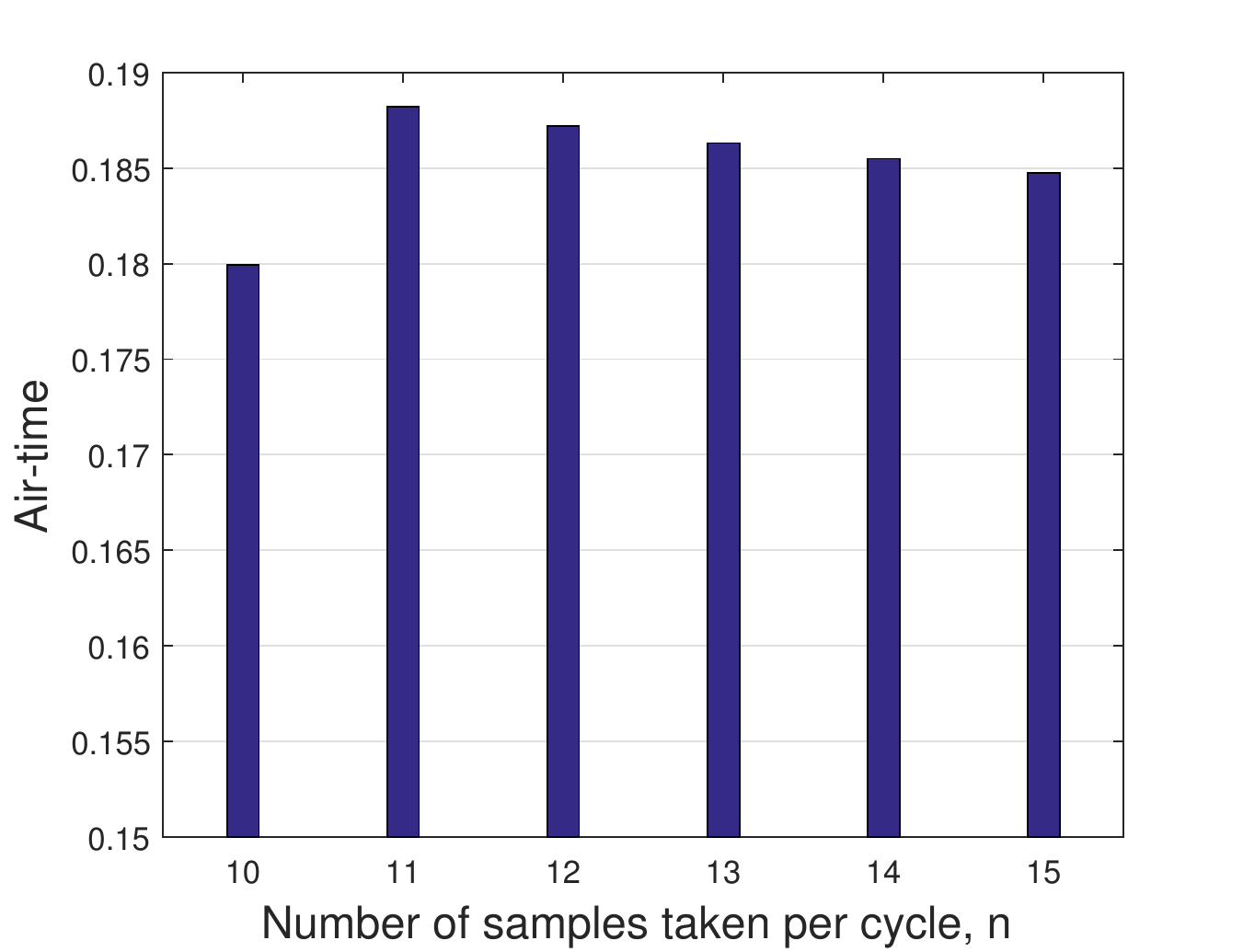}\label{Fig:airtimeE1}
}
  \caption{Energy consumption and air-time vs the maximum number of samples allowed per cycle. Parameter values are listed in Table~\ref{Exmp1Parameters}.}
\end{figure}

Fig.~\ref{Fig:energyE1} plots the energy consumed by each node and the energy received from the H-AP per cycle. It can be seen that the node with the smallest number of samples receives the least amount of energy and the consumed energy equals to the received amount. The other 5 nodes consume less than the amount that has been charged. Fig.~\ref{Fig:airtimeE1} shows the corresponding air-times for the 6 nodes. \emph{Air-time} is defined as the fraction of time used for transmissions from node $i$, including successful and collided transmissions~\cite{PFcoding}, given by
\begin{equation*}
t_i=\frac{P_i^{succ}T_i^{succ}+P_i^{col}T^{col}}{P^{idle}\sigma+\sum\limits_{j=0}^{N-1}P_j^{succ}T_j^{succ}+P^{col}T^{col}}
\end{equation*}
 It can be seen that if energy is not a constraint when determining the node attempt probability $\tau_i$, the optimized solution assigns a lower air-time to a node with a longer packet to send (i.e. a larger $n_i$). This is to be expected as the proportional fairness design is to bring down the throughput of more aggressive nodes while giving more transmission opportunities to less aggressive ones. However, if the determination of node transmission attempt rate is constrained by the amount of energy available, in this example this is the case for the node with $\bar{n}_i=10$,  the air-time is brought down to guarantee that the consumed energy does not exceed the available amount.

\subsection{Impact of distance from the H-AP}

The RF power received by end nodes is mainly determined by the distance from the power transmitter.  Devices far from the H-AP receive less wireless energy than nearer nodes, but need to transmit data with a greater power level at a lower PHY data rate. This is the so called doubly near-far problem~\cite{276}. In this example, we consider a scenario that nodes have distinct levels of received RF power, and accordingly distinct data transmitting and receiving powers and PHY rates. A network consisting of 6 nodes is considered.  The received RF power,  data receiving power and PHY rate for the 6 nodes are listed in Table~\ref{Exmp2Parameters1}. The ratio between the transmitting power and the receiving power is fixed to be 1.32. Other parameters are assumed to be the same, as listed in Table~\ref{Exmp2Parameters2}. 

\begin{figure}
\centering \subfigure[Energy consumed and received]{
  \includegraphics[width=0.46\columnwidth, height=3cm]{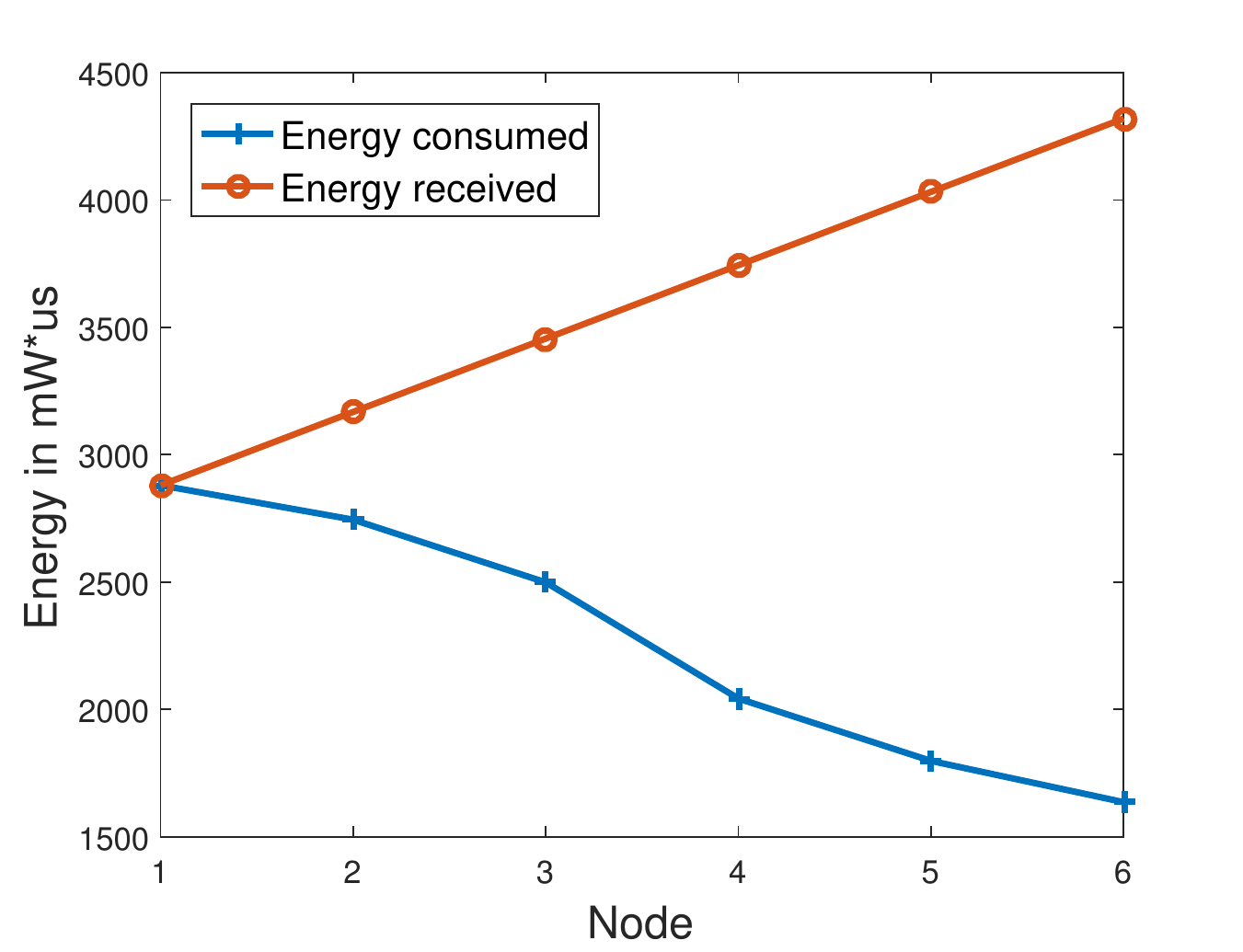}\label{Fig:energyE2}
 }
 \subfigure[Air-time]{
   \includegraphics[width=0.46\columnwidth, height=3cm]{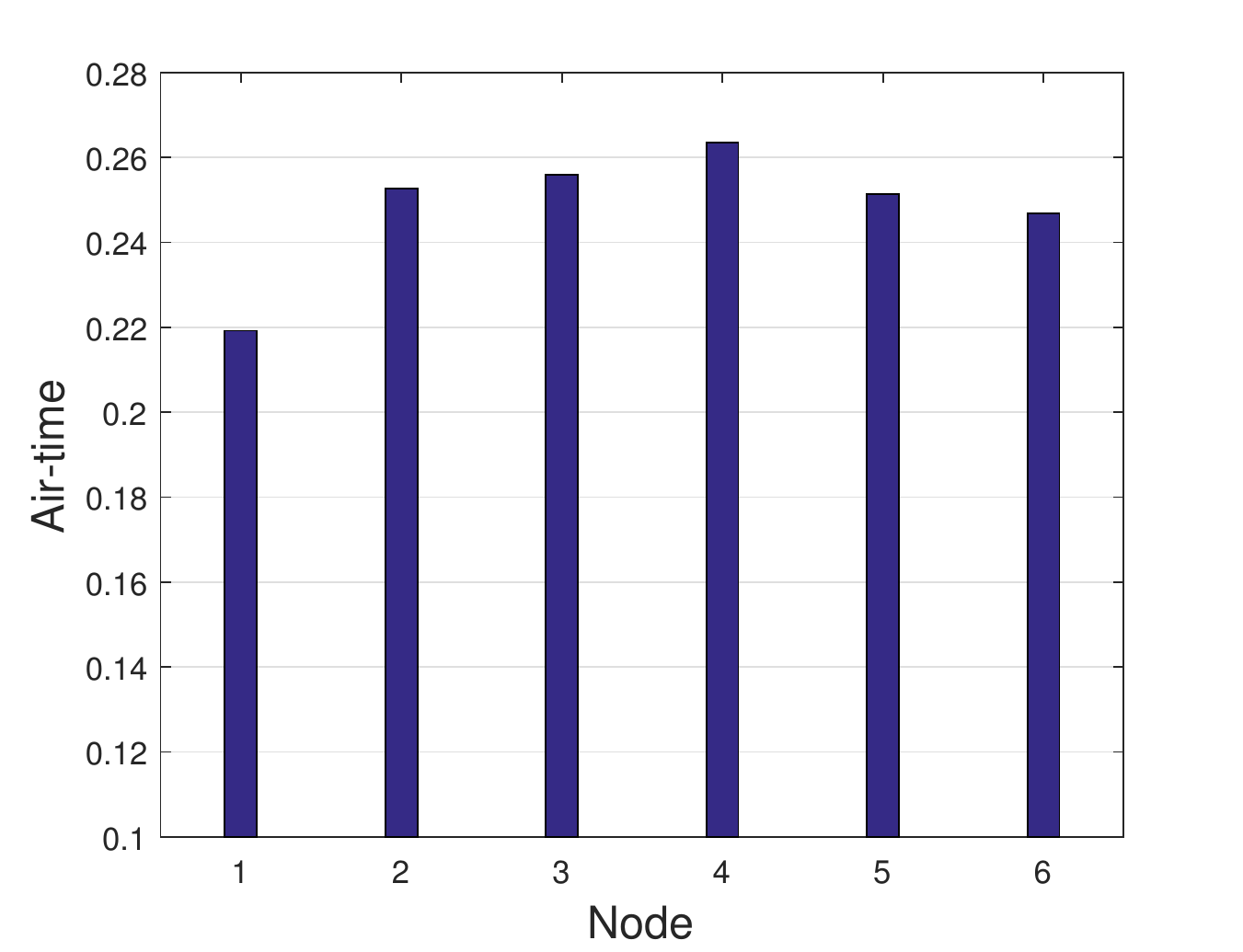}\label{Fig:airtimeE2}
}
  \caption{Energy consumption and airtime vs distinct distances from the H-AP. Parameter values are listed in Table~\ref{Exmp2Parameters1} and~\ref{Exmp2Parameters2}.}
\end{figure}

The optimized solution indicates that for each node the optimal number of samples taken per cycle is the given upper bound, the same as observed in the previous example.  As the farthest node receives the least amount of wireless power but consumes the largest amount of energy for data communication, the optimized strategy is to satisfy the energy consumption need for the farthest node, as shown in Fig.~\ref{Fig:energyE2}. All other nodes receive more power than they actually consume.  Fig.~\ref{Fig:airtimeE2} plots the corresponding air-times. It can be seen that for nodes 2-6 which have sufficient energy to consume, as the PHY rate increases, the air-time increases up to node 4 and then drops down from node 5. This shows how the proportionally fair allocation tackles the doubly near-far unfairness. Although node 4 is farther from the H-AP and receives less RF energy from H-AP compared to node 5 and 6, the optimized strategy allocates a longer air-time to node 4 to achieve the trade-off between the sum-throughput and the fairness across nodes. But for node 1-3, the near-far problem still exists     simply because the objective in this work is to proportionally maximize the sum-throughput, similar to other sum-rate maximization work, if the contribution to the sum-throughput is not significant enough, the fairness cannot be guaranteed. 
\begin{table}\tiny
\caption{Distinct parameter values for evaluating the impact of distance from the H-AP }\label{Exmp2Parameters1}
\centering
\begin{tabular}{|c|c|c|c|c|c|c|}
  \hline
  Node   & 1& 2&3&4&5&6     \\ \hline
    Received RF power (mW) & 10&11&12&13&14&15\\ \hline
 Data receiving power (mW) & 15&14&13&12&11&10\\ \hline
 PHY data rate (Mbps) & 5.5&5.5&6&9&11&12\\ \hline
\end{tabular}
\end{table}

\begin{table}\tiny
\caption{Same parameter values for evaluating the impact of distance from the H-AP}\label{Exmp2Parameters2}
\centering
\begin{tabular}{|c|c|c|c|c|c|c|c|}
  \hline
    $\bar{n}_i$ & 10 & $P_i^{L}$   & 9mW & $l_i$         & 10 bytes & $h_i$         & 3      \\ \hline
   $P_i^{A}$      &    5mW  &   $P_i^{P}$            &    6mW  &  $g_i$         & 2 & & \\ \hline
\end{tabular}
\end{table}

\section{Conclusions}
This paper considers the deployment of a H-AP in an 802.11-based wireless powered IoT network. The proportionally fair allocation of throughputs across IoT nodes is considered under the constraints of energy neutrality and CPU capability for each device. The joint optimization of wireless powering duration and data transmission airtime is solved by using DC programming and block coordinate descent algorithms. The optimized solution suggests that the number of samples an IoT node should take in each cycle should equal to the maximum number that its CPU can process concurrently, and the individual throughput across nodes can be effectively balanced by tuning the air-time in terms of energy constraint. Simulations have verified our analysis.

\bibliography{Ref}

%

\end{document}